\documentclass[]{raa}            
\usepackage{graphicx,times}
\usepackage{natbib}

\providecommand{\bm}[1]{\mbox{\boldmath$#1$\unboldmath}}

\begin{document}
   \title{On the possibility of disk-fed formation in supergiant high-mass X-ray binaries}

 \volnopage{ {\bf 2018} Vol.\ {\bf X} No. {\bf XX}, 000--000}
   \setcounter{page}{1}

   \author{Ali Taani\inst{1},  Shigeyuki Karino\inst{2}, Liming Song\inst{3}, Mashhoor Al-Wardat\inst{4},
   Awni Khasawneh\inst{5} and Mohammad K. Mardini\inst{6,7}      }

 \institute{Physics Department, Faculty of Science, Al-Balqa Applied University, 19117 Salt, Jordan {\it ali.taani@bau.edu.jo}\\
\and
Faculty of Science and Engineering, Kyushu Sangyo University, 2-3-1 Matsukadai, Higashi-ku, Fukuoka 813-8503, Japan\\
\and
Key Laboratory of Particle Astrophysics, Institute of High Energy Physics, Chinese Academy of Sciences, Beijing
100049, China\\
\and
Department of Physics, Al al-Bayt University, Mafraq 25113, Jordan\\
\and
Royal Jordanian Geographic Center, Amman, 11941 Jordan\\
\and
Key Lab of Optical Astronomy, National Astronomical Observatories, Chinese Academy of Sciences, Beijing 100012, China
\and
University of Chinese Academy of Sciences, Beijing 100049, China\\
\vs \no
   {\small accepted  1 September  2018}
}


\abstract{We have considered the existence of neutron star magnetic field given by
the cyclotron lines.
We collected the data of 9 sources of high-mass X-ray binaries with
supergiant companions as a case of testing  our model, to demonstrate
their distribution and evolution.
The wind velocity, spin period and magnetic field strength are studied
under different mass loss rate.
In our model, correlations between mass-loss rate and wind velocity are
found and can be tested in further observations.
We examined the parameter space where wind accretion is allowed,
avoiding barrier of rotating magnetic fields, with robust data of
magnetic field of neutron stars.
Our model shows that most of sources (6 of 9 systems) can be fed by the wind with
relatively slow velocity, and this result is consistent with previous
predictions. In a few sources, our model cannot fit under the standard wind
accretion scenario.
In these peculiar cases, other scenarios (disk formation, partial Roche
lobe overflow)  should be considered.
This would provide information about the evolutionary tracks of various
types of binaries, and thus show a clear dichotomy behavior in wind-fed
X-ray binary systems.
}
   \authorrunning{Taani et al. }            
   \titlerunning{On the possibility of disk-fed formation in supergiant high-mass X-ray binaries} 
   \maketitle


%
\section{Introduction}
\label{sect:intro}

High Mass X-ray Binaries (HMXBs) with inferred magnetic
fields  just on the surface of compact companion $B\sim 10^{12}$ G, are composed of two subclasses: Be
X-ray binaries ($\sim \%~80$) in the Galaxy (e.g. van den Heuvel 2004; Liu et al. 2006; Taani 2016), with orbital periods
ranging from $\sim $15 days to several years, and with relatively
low mass companions ($\rm \sim8 ~to ~20 \rm M_{\odot}$). In contrast, the second group of HMXBs is the supergiant (SG) sources. They consist of OB SG mass donors ($\rm \sim18 ~to ~ over~ 40 \rm M_{\odot}$) and a compact object accreting from the strong stellar wind with short orbital periods ($ \leq $ 11 days) like Cen X-1 and Vela X-1, and only about a dozen known members (see e.g. Bhattacharya \& van den Heuvel 1991; Reig et al. 2009; Taani et al. 2012a,b; Walter et al. 2015; Taani \&  Khasawaneh 2017; Dai et al. 2017).

The cyclotron lines are detected as absorption lines in
high-energy spectra (see Tr\"{u}emper et al. 1978, for full details and references) of magnetized accreting neutron stars (NSs). They form
in the presence of a strong magnetic field due to resonant
scattering processes with electrons (Voges et al.
1982; Wilson et al. 2008) in order to provide the only direct
estimate of the magnetic field strength of an accreting NSs. The presence of cyclotron feature has been reported for SG-HMXB listed in our sample at $\rm \geq 17$ keV (see table 1). There are also factors of several uncertainties on the value of the NS magnetic field derived from
these lines due to unknowns in the geometry and emission mechanisms of the system
(see, e.g., Reig et al. 2009; Mushtukov et al. 2015; Walter et al. 2015). Nishimura (2003) demonstrated that the variations on the
strength of a magnetic field affect the profile of the cyclotron line.
It is noteworthy to mention here, that most observed cyclotron lines have been detected above 10 keV  (e.g. Makishima et al. 1990;  Cusumano et al. 1998; Coburn et al. 2002) are interpreted as electron features, with inferred magnetic fields $B\sim 10^{12}$ G. However the NS magnetic field of HMXB may decay a little by accreting matter  (see, e.g. Zhang \& Kijima 2006 and references therein).

As the mass-loss rate through stellar winds may influence on the stellar evolution
and modifies stellar spectrum (Krti\v{c}ka et al. 2016). Many observational methods can be used to derive wind mass-loss rate.
This will also allow to distinguish the phenomenology of the X-ray sources and their optical counterpart in a natural way (Wei et al. 2010; Shakura et al. 2012; Saladino et al. 2018; Taani et al. 2018), such as the shape of X-ray emission lines in cool wind (MacFarlane et al. 1991; Feldmeier et al. 2003),
the strength of the UV P Cygni lines (Lamers et al. 1999), H $\alpha$ emission line (Puls at al. 2006) and near-infrared emission lines (Najarro et al. 2011).

Several disk-fed models  have been proposed to explain the   disk accretion mechanism, including the  circumstellar disk (see i.e. Wang 1995; Dai \& Li 2006; Klu\'{z}niak \& Rappaport 2007;  Shakura et al. 2012; Taani et al. 2018). For a recent review of stellar wind models, we refer to Krti\v{c}ka et al. (2016) and Krti\v{c}ka \& Kubat (2017).
One of the most accepted model about the stellar wind properties, is the quasi-spherical accretion that captured matter from stellar wind of the star (Shakura et al. 2012; Postnov et al. 2017).
In this work, we have concentrated more on the study
and interpretation of the evolution of the  wind accretion velocity, spin period, magnetic field, mass-loss rate for 9 HMXBs with SG companions. In addition, we study the disk-fed NS formation through the relations between mass-loss rate and wind velocity.
The article is organized as follows: Sect. 2, deals with consideration of our model. Sect. 3, discusses accretion regimes. The obtained results
on $\dot{m}_{\rm{w}}$, $v_{\infty}$, B$_{field}$,  and P$_{spin}$ for several SG-HMXB listed in our sample are discussed in Sec. 4 and conclusions are summarized in Sec. 5.



\section{The wind model}


In  order  to  build  an  evolutionary framework for HMXBs, it is essential to map the mass-loss processes  during
the various evolutionary stages.
According to eq.(2) in Karino \& Miller 2016 (KM16), we apply the equilibrium period equation (the magnetospheric
radius is equal to the NS corotation radius, $r_{\rm{m}}$=$r_{\rm{co}}$), where the disk accretion is assumed (Bhattacharya \& van den Heuvel 1991). It is determined by the long term averaged mass accretion rate
(Tong 2015), thus it is convenient for the wind accretion calculation. The magnetic field strength of NS can be estimated by
\begin{equation}
B_{\rm{NS}} = 2.184 \times 10^{12} \rm{G} \times \zeta^{1/2}
\left( \frac{\dot{M}}{10^{18} \rm{g\, s}^{-1}} \right)^{1/2}
\left( \frac{P_{\rm{s}}}{1 \rm{s}} \right)^{7/6} ,
\end{equation}
where we assume that the mass of NS is 1.4 $M_\odot $, and radius of NS is 10~km.
For the wind accretion ($\zeta \approx 1$) or disk accretion ($\zeta \approx 0.1$).

Since the mass-loss rate is one of the key parameters that determines the influence of the stellar wind
on stellar evolution and on the circumstellar medium (Krti\v{c}ka \& Kub\'{a}t  2017). We assume the
Bondi-Hoyle-Littleton accretion (Bondi \& Hoyle 1944) in
 a smooth wind, to estimate the mass accretion rate onto the NS as

\begin{equation}
 \dot{M}_{\rm{acc}} = \rho_{w} R_{\rm{acc}}^{2} v_{\rm{rel}},
\end{equation}
where $\rho_{w}$ is the density of the wind during the steady state of
spherical wind.
$v_{\rm{rel}}$ is the relative velocity of the wind:

\begin{equation}
v_{\rm{rel}} = \left( v_{\rm{orb}}^2 + v_{\rm{w}}^2 \right)^{1/2},
\end{equation}
where $v_{\rm{orb}}$ and $v_{\rm{w}}$ denote the orbital velocity and the wind velocity, respectively.
Note that, the stellar wind velocity ($v_{\rm{w}}$) is usually larger (typically $100 - 1000 \rm{km \, s}^{-1}$)
than the orbital velocity $v_{\rm{orb}}$ for typical wind-fed systems with $P_{\rm{orb}} \approx 10 \rm{d}$ (Shakura et al. 2012).
Hence, in these systems, we could neglect the $v_{\rm{orb}}$ from the calculation. $R_{\rm{acc}}$ is the accretion
radius (Bondi radius) defined by
\begin{equation}
 R_{\rm{acc}} = \frac{ 2 G M_{\rm{NS}} }{ v_{\rm{rel}}^2} .
\end{equation}
We adopt the standard Castor et al. (1975) formula for the wind velocity $v_{\rm{w}}$, which assumed
a stationary, homogeneous, and spherically symmetric outflow,
\begin{equation}
v_{\rm{w}} = v_{\infty} \left( 1 - \frac{R_{\rm{d}}}{a} \right)^{\beta}
\label{eq:vwind}
\end{equation}
In this study, we assume $\beta$ which is a free input parameter to be $\beta = 1$ (Puls et al. 2006).
$R_{\rm{d}}$ is the radius of the donor.
$v_{\infty}$ denotes the terminal velocity of the wind.
The density of the wind in the stellar atmosphere can be derived from the continuity equation
\begin{equation}
\dot{M}_{\rm{w}} = 4 \pi a^{2} \rho_{\rm{w}} v_{\rm{w}}.
\end{equation}
Here $\dot{M}_{\rm{w}}$ is the mass loss rate from the donor, and $a$ denotes the orbital radius of the
system. (Here we assume the circular orbit.)
By combining equations (2)-(6), we got
\begin{equation}
 \dot{M}_{\rm{acc}} = \left( \frac{G M_{\rm{NS}}}{a} \right)^{2}
\frac{\dot{M}_{\rm{w}}}{\pi v_{\rm{rel}}^{3} v_{\rm{w}}}.
\end{equation}
The mass accretion rate can be derived from the X-ray luminosity as follows,

\begin{equation}
L_{\rm{X}} \simeq \frac{G M_{\rm{NS}} \dot{M}_{\rm{acc}}}{R_{\rm{NS}}}.
\end{equation}


%
Hence, $\dot{M}_{\rm{acc}}$ could be known from observed luminosities.
Adopting formulae by the Vink et al. (2001), we can estimate $\dot{M}_{\rm{w}}$ from the parameters of donors;
most of SG-type donors in our sample set shows $\dot{M}_{\rm{w}} \sim 10^{-7} - 10^{-6} M_{\odot} \rm{y}^{-1}$.





Assuming that the NS has a typical parameters ($M_{\rm{NS}} = 1.4 M_{\odot}$ and $R_{\rm{NS}} = 10 \rm{km}$),
the orbital parameters and wind parameters can be obtained from the donor mass and radius.
Adopting the donor parameters given by previous studies such as Coley et al. (2015), Falanga et al. (2015),
Rawls et al. (2011) and Reig et al. (2016),  we can obtain the possible parameter range of the unknown wind parameters
(terminal velocity $v_{\infty}$ of the wind and the mass loss rate $\dot{M}_{\rm{w}}$ of the donor) with the robust values
of the B-field obtained from cyclotron lines (see Table~1).

\section{Accretion regime}

In Stella et al. (1986) and Bozzo et al. (2008), they considered three typical radii:
accretion radius ($r_{\rm{a}}$), magnetic radius ($r_{\rm{m}}$) and corotation radius ($r_{\rm{co}}$)) and
divided the parameter space into 5 accretion regimes based on their magnitude relation.
That is, the parameter space can be categorized into
(A)supersonic inhibition regime ($r_{\rm{m}} > r_{\rm{a}}, r_{\rm{co}}$),
(B)subsonic inhibition regime ($r_{\rm{co}} > r_{\rm{m}} > r_{\rm{a}}$),
(C)supersonic propeller regime ($r_{\rm{a}} > r_{\rm{m}} > r_{\rm{co}}$),
(D)subsonic propeller regime ($r_{\rm{co}} , r_{\rm{a}} > r_{\rm{m}}$ , $\dot{M} < \dot{M}_{\rm{c}}$),
where $\dot{M}_{\rm{c}}$ denotes the critical limit where radiative cooling starts working (see Bozzo et al. 2008)
and (E)direct accretion regime.
These radii, in turn, depend on: $\dot{M}_{\rm{w}}$, $v_{\infty}$, $B_{\rm{NS}}$, and $P_{\rm{spin}}$.
Among these parameters, now we have robust data of $P_{\rm{spin}}$ given by light curve analysis and $B_{\rm{NS}}$
given by cyclotron-feature observation.
We have performed the above segmentation in the magnetic field-wind velocity space intended for several SG-HMXB.
Although the possible effect of  angular momentum loss during the propeller regime still has to be  extensively
discussed (see, e.g., Pringle \& Rees 1972; Wang \& Robertson 1985; Shakura et al. 2012; Dai et al. 2016),
in this study we concentrate our attention on the direct accretion regime only.

The graphical illustration of the accretion regime is given in Fig. 1: the different accretion regimes (A) to (E) are divided
by dashed lines.
The shaded region denotes the direct accretion regime; only HMXBs in this regime can be observed as bright X-ray sources.
The horizontal solid lines denote the wind velocity corresponding to $v_{\inf} = 1,500 \rm{km \, s}^{-1}$ (upper line) and
$700 \rm{km \, s}^{-1}$ (lower line), respectively.
(From recent observations, it is suggested that the wind velocity in persistent SG HMXBs are relatively slow; for example,
in Vela X-1, $v_{\infty} = 700 \rm{km \, s}^{-1}$), while the wind velocity in supergiant fast X-ray transients (SFXT) systems seems as fast as
$v_{\infty} = 1,500 \rm{km \, s}^{-1}$.)
Tentatively, we fix the mass loss rate of the donor as $\dot{M}_{\rm{w}} = 5 \times 10^{-7} M_{\odot} \rm{y}^{-1}$,
so as to show the typical results.

The vertical line is the magnetic field strength which is given by observed cyclotron lines.
In the case that the cross-section of the vertical line and the horizontal band between two horizontal lines comes
in the shaded region, the system can be considered a {\it normal} wind-fed HMXB, which can be understood only with
the previous model.
Actually, most of our sample shows good agreement the standard scenario.
However, in some systems, we need to reconsider our models in order to explain the fact that such crossing points
come out of direct accretion regime.

Fig. 1 shows the results of our sample. 6 of 9 sources (4U1907, 4U1538, J16493, 2S0114, J16393 and J18027) are found to be in good agreement with the standard wind-fed scenario as described above.
A note should be made concerning the Vela X-1, because this source can be satisfied our constraint if the mass-loss rate reaches $\dot{M} = 5 \times 10^{-7}M_{\odot} \rm{y}^{-1}$. While if this value is larger, however, this system cannot satisfy the above condition when $v_{\infty}$ is large.
Assuming that the mass-loss rate is reasonably high ($\dot{M} \sim 10^{-6} M_{\odot} \rm{y}^{-1}$), the wind velocity
would be limited below $1,000 \rm{km \, s}^{-1}$.
This result is consistent with recent result given by Gim\'{e}nez-Garc\'{\i}a et al. (2016).
On the other hand, 2 systems (LMC X-4 and OAO1657) cannot satisfy the wind-fed condition.
That is, the reasonable band-region of $v_{\infty}$ cannot cross with the observed B$\rm-_{field}$ in the direct accretion
regime (shaded area in Fig.~1).

Since LMC X-4 is one of the tightest X-ray binaries with SG donor, it has been argued that Roche lobe of the donor
is filled in this system.
If it is true, in this system the accretion matter is transfered via Roche lobe overflow scenario passing L1 point,
and the wind-fed scenario cannot be applied.
In recent study, however, it is suggested that the donor in LMC X-4 is much smaller than its Roche radius
(Falanga et al. 2015).
Another possibility is that our assumption of the mass-loss rate ($\dot{M} = 5 \times 10^{-7} M_{\odot} \rm{y}^{-1}$)
is not valid.
To check this validity, we vary the mass-loss rate and try the computations again.
In Fig.~2, the same expressions are shown for LMC X-4 in the cases of high and low mass-loss rate:
$5 \times 10^{-8} M_{\odot} \rm{y}^{-1}$ (left panel) and
$5 \times 10^{-8} M_{\odot} \rm{y}^{-1}$ (right panel), respectively.
When the mass-loss rate is quite large and the wind velocity is slow, the wind-fed condition could be satisfied.
However, the donor of LMC X-4 is not highly evolved if it is not filled its Roche lobe, and it is questionable that
such a donor emit such a dense and slow wind.
Hence, in this system, the intermediate state between Roche lobe overflow and wind-fed accretion
(Nagae et al. 2004) may be realized.

OAO1657 also cannot be explained by standard wind-fed scenario in Fig.~1.
Also for this source, we try to compute the different mass-loss rate and show the results in Fig.~3.
From this figure, we can see that if the mass-loss rate is extremely high (and wind velocity is slow preferably),
the system can be satisfy the wind-fed condition.
In this system, the donor may be WR star and may emit very dense wind.
Therefore, such a dense and slow wind may be possible in this system.
However, if the donor is the WR star, its evolutionary pass becomes another puzzle (Mason et al. 2012).
In this source, further studies both in theoretical side and in observational way are required.

\newcounter{rit} 
\newcommand{\rit}{\refstepcounter{rit}\therit}

\begin{table*}
\caption{Parameters of HMXBs with supergiant companions founded with their the cyclotron lines.} \label{O-I}
\setlength{\tabcolsep}{2pt}
\renewcommand{\arraystretch}{0.1}
\begin{tabular}{lcccccccccl}
\hline \hline \noalign{\smallskip}
 \multicolumn{1}{c}{Object} &
\multicolumn{1}{c}{$P_{\rm spin}$} & \multicolumn{1}{c}{$P_{\rm
orbit}$} & \multicolumn{1}{c}{$e$} &\multicolumn{1}{c}{$distance$} & \multicolumn{1}{c}{$E_{\rm cyc}$}
&\multicolumn{1}{c}{$M_{\rm NS}$} &\multicolumn{1}{c}{$M_{\rm comp.}$} &
\multicolumn{1}{c}{Ref.} \\
\multicolumn{1}{c}{}& \multicolumn{1}{c}{(s)} & \multicolumn{1}{c}{(d)} & \multicolumn{1}{c}{} &\multicolumn{1}{c}{(kpc)} &\multicolumn{1}{c}{(keV)} &
 \multicolumn{1}{c}{$(\rm M_{\odot})$} &
\multicolumn{1}{c}{$(\rm M_{\odot})$} &
\multicolumn{1}{c}{}\\
\hline \noalign{\smallskip}
\\ 
4U 1907$+$09 & 439 & 8.37 & 0.28&5&18.8$\pm$0.4  &1.4&27   &           \ref{1998Cusumano}, 
              \ref{2002Coburn}, \ref{2010Rivers}\\
4U 1538$-$52 & 529 & 3.73&0.18&4.5             & 21.4$^{+0.9}_{-2.4}$  &1.06&16.4&
             \ref{2002Coburn}, \ref{1990Clark}, 
              \ref{2001Robba}, \ref{2009Rodes-Roca} \\
Vela X-1 & 283 & 8.96 & 0.09&1.4&54$^{+0.5}_{-1.1}$ &1.86&23.8&
             \ref{1996Kretschmar}, \ref{1999Makishima}, \ref{2002Kreykenbohm}, \ref{2007Schanne}, \ref{2015Walter}\\
Cen X-3 & 4.8 & 2.09 & 0.01&5.7&30.4$^{+0.3}_{-0.4}$ &   1.5&20&     \ref{2002Coburn}, \ref{1999Makishima}, \ref{2015Walter}, \ref{1998Santangelo}  \\
LMC X-4 & 13.5 & 1.4 & 0.06&50&100$\pm$2.1 &1.25&14.5&
         \ref{1999Makishima}, \ref{2001LaBarbera} \\
OAO1657-415 & 37.7& 10.4& 0.1&7.1&36&1.42&41&
 \ref{1999Orlandini}, \ref{2010Denis}, \ref{2011Pottschmidt}\\
J16493-4348 & 1069 & 6.78 & 0.25&10.7&33$\pm$4  &--$^{\dag}$&--&  \ref{2006Bodaghee}, \ref{2010Nespoli}, \ref{2011D'Ai},\ref{2016Bodaghee}  \\
2S 0114+65& 9700& 11.6& 0.16&7.2&22& 1.7&16&  
\ref{2005Bonning}, \ref{2006denHartog}\\
J18027-201&140&4.6&0.2&12.4&23&1.6&21.8 &\ref{2011Mason}, \ref{2017Lutovinov}\\
\hline \noalign{\smallskip}
\end{tabular}

$^{\dag}$The X-ray mass function is $6.5\pm1.1\rm{M}_{\odot}$ (see Thompson et al. 2006, for full details)\\
\textbf{References:}
(\rit\label{1998Cusumano}) Cusumano et al. 1998;
 (\rit\label{2002Coburn}) Coburn et al. 2002;
 (\rit\label{2010Rivers}) Rivers et al. 2010;
 (\rit\label{1990Clark}) Clark et al. 1990;
 (\rit\label{2001Robba}) Robba et al. 2001;
  (\rit\label{2009Rodes-Roca}) Rodes-Roca et al 2009;
 (\rit\label{1996Kretschmar}) Kretschmar et al. 2005;
  (\rit\label{1999Makishima}) Makishima et al. 1999;
  (\rit\label{2002Kreykenbohm}) Kreykenbohm et al. 2002
 (\rit\label{2007Schanne}) Schanne  et al. 2007
     (\rit\label{2015Walter}) Walter et al. 2015;
   (\rit\label{1998Santangelo}) Santangelo et al. 1998;
(\rit\label{2001LaBarbera}) Barbera et al. 2001;
(\rit\label{1999Orlandini}) Orlandini et al. 1999;
(\rit\label{2010Denis}) Denis et al. 2010;
(\rit\label{2011Pottschmidt}) {Pottschmidt} et al. 2011
(\rit\label{2006Bodaghee}) {Bodaghee} et al. 2006
  (\rit\label{2010Nespoli}) {Nespoli} et al. 2010
  (\rit\label{2011D'Ai}) {D'Ai} et al. 2011
  (\rit\label{2016Bodaghee}) {Bodaghee} et al. 2016.~
 (\rit\label{2005Bonning}) {Bonning} et al. 2005
(\rit\label{2006denHartog}) {denHartog} et al. 2006
(\rit\label{2011Mason}) {Mason} et al. 2011.~
(\rit\label{2017Lutovinov}) {Lutovinov} et al. 2017.~
\end{table*}

%
%

\begin{figure}
\begin{center}
\includegraphics[width=4.7cm]{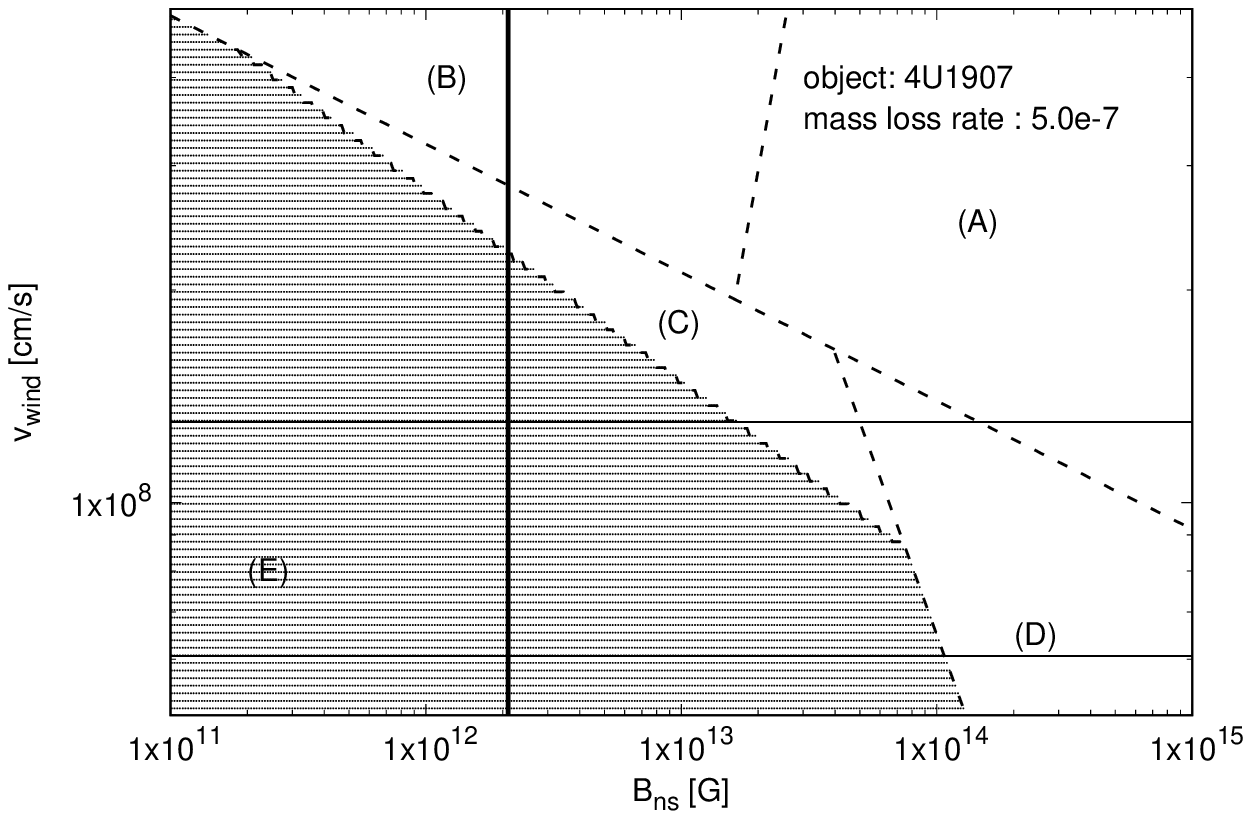}
\includegraphics[width=4.7cm]{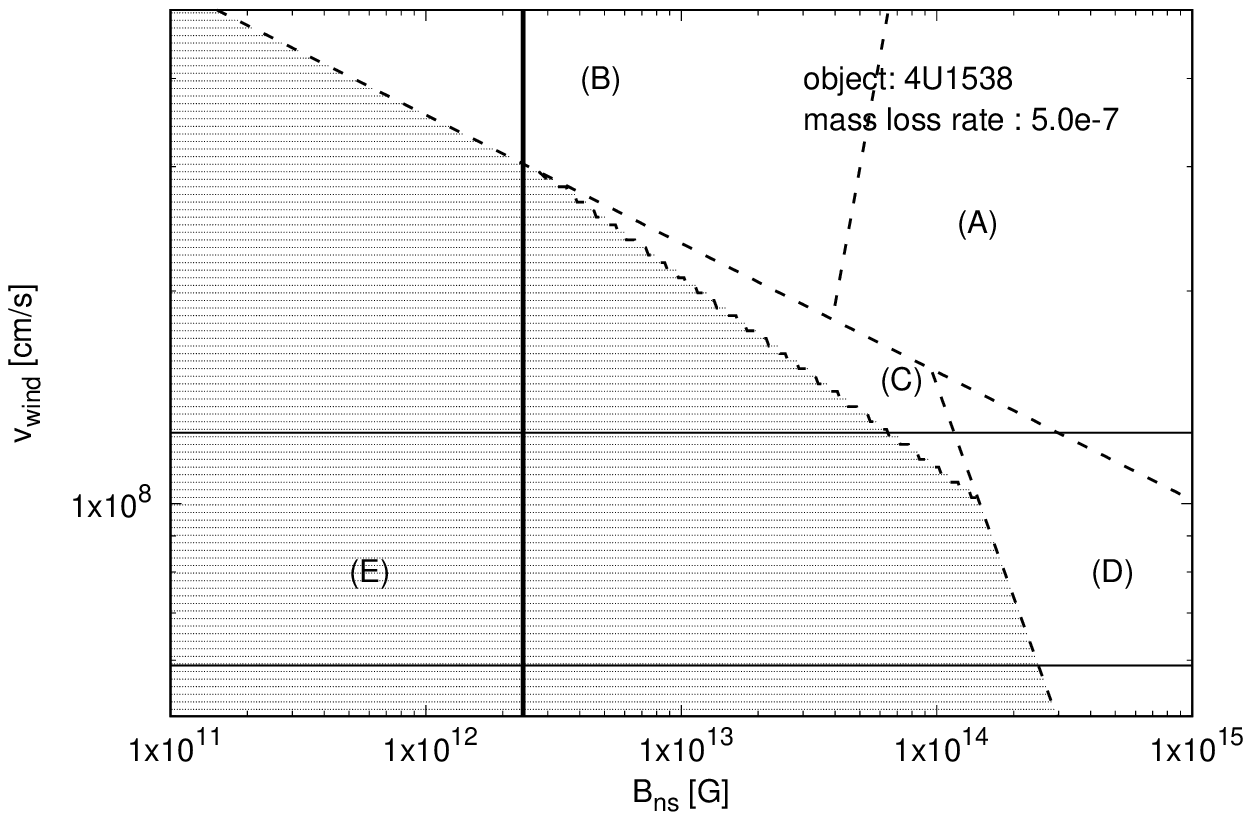}
\includegraphics[width=4.7cm]{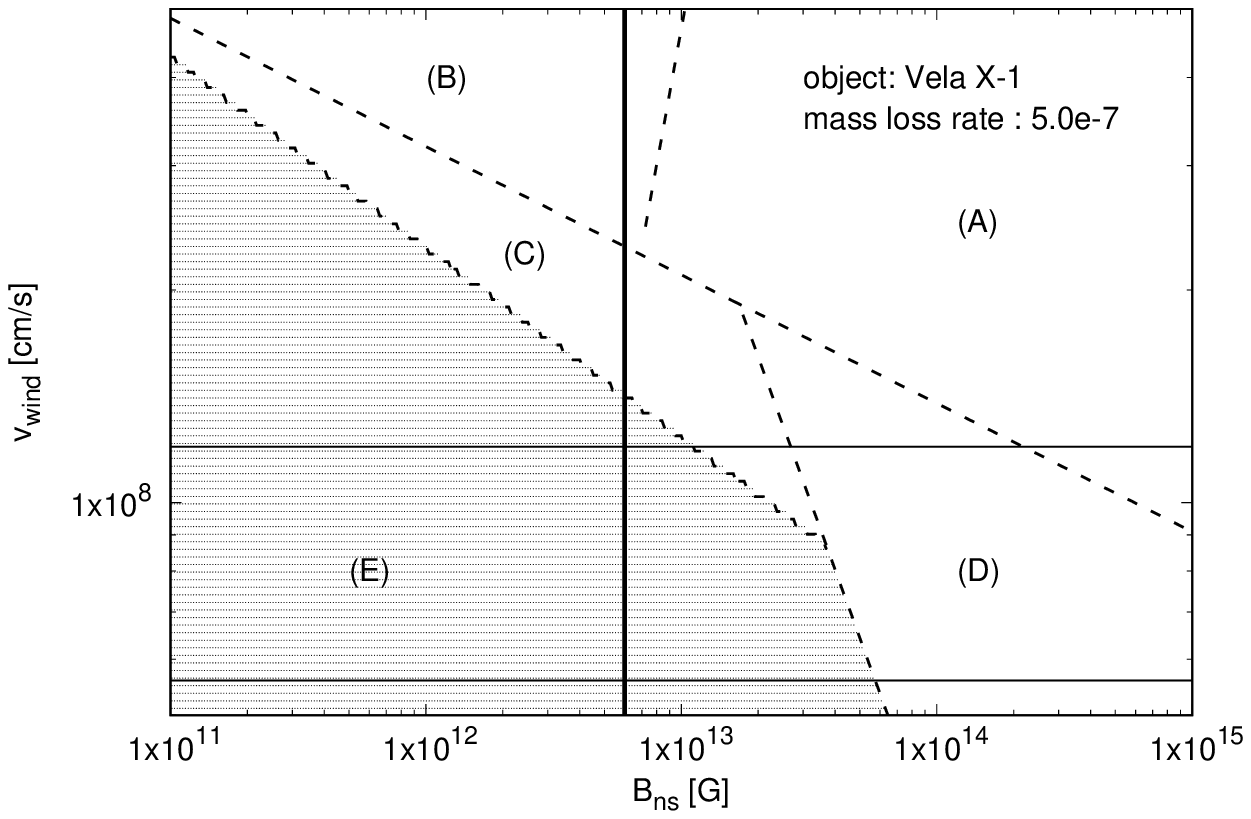} \\
\includegraphics[width=4.7cm]{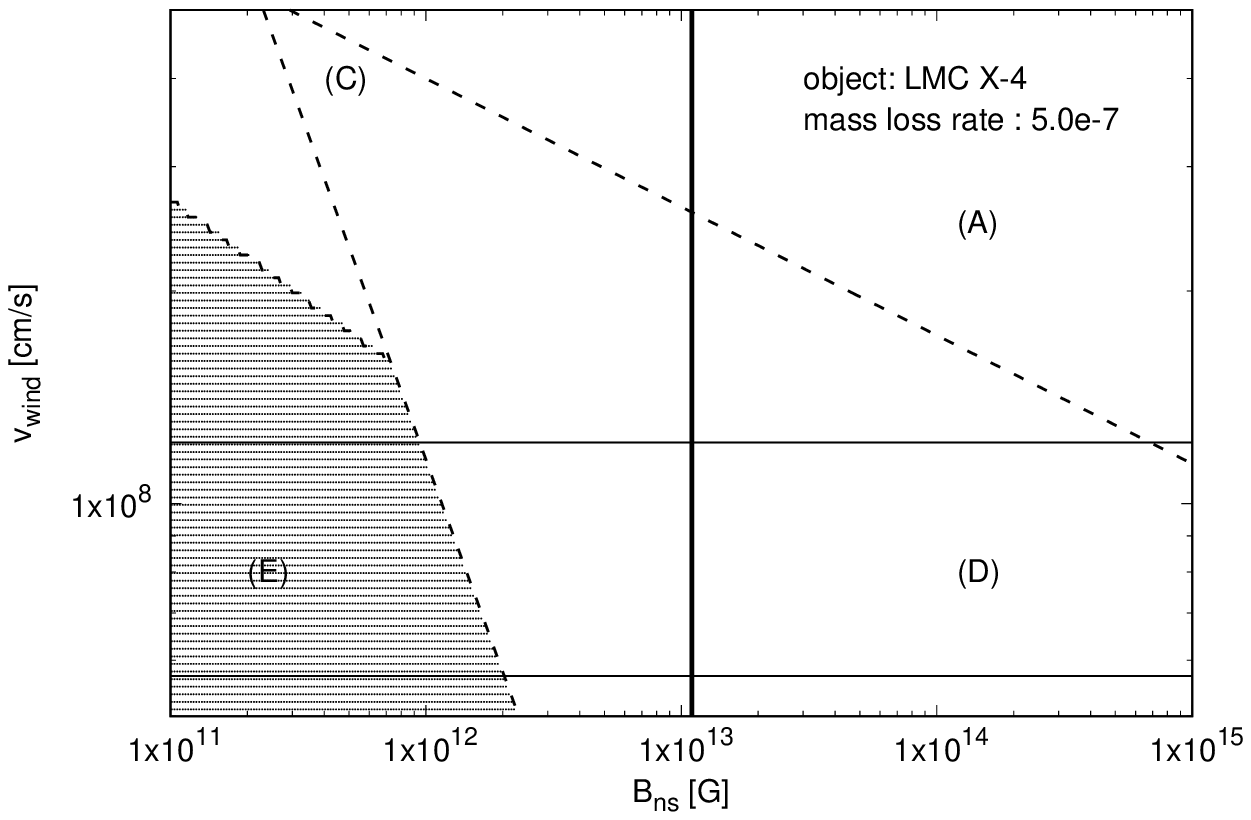}
\includegraphics[width=4.7cm]{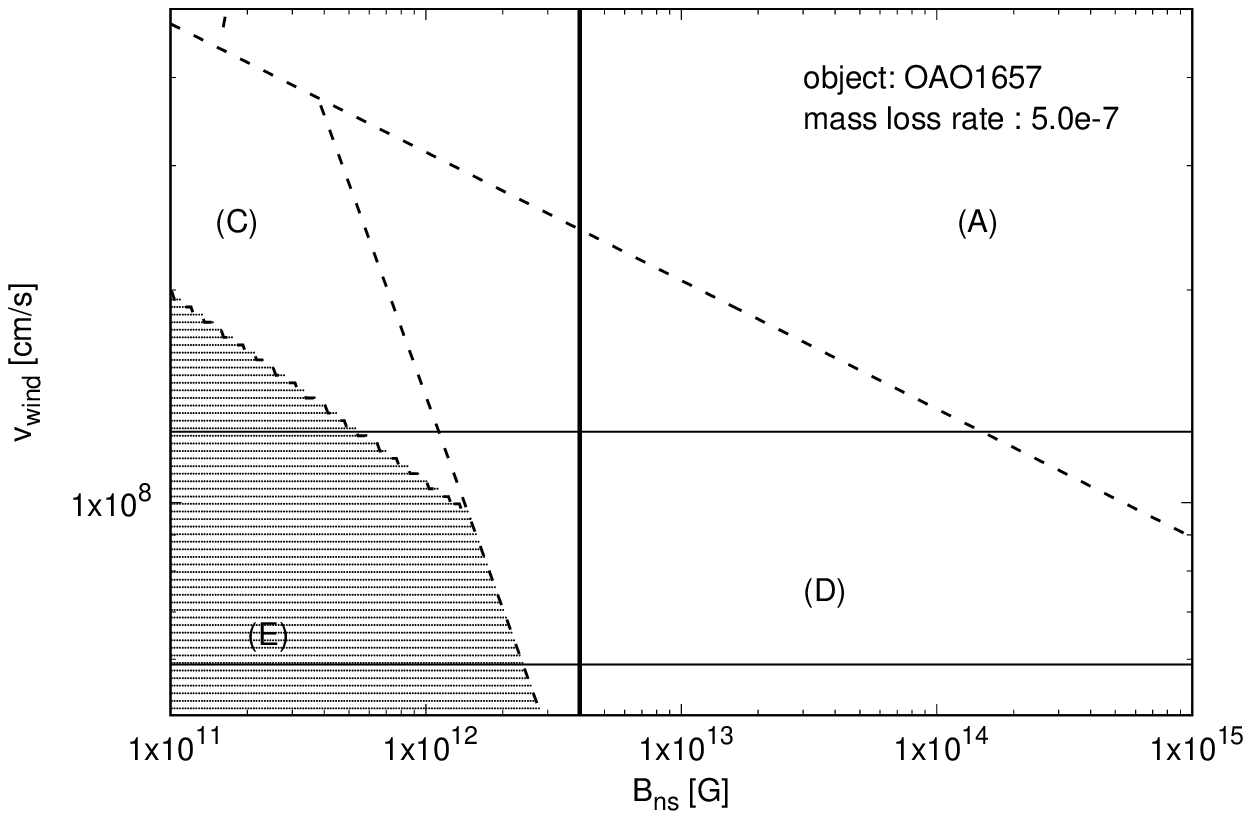}
\includegraphics[width=4.7cm]{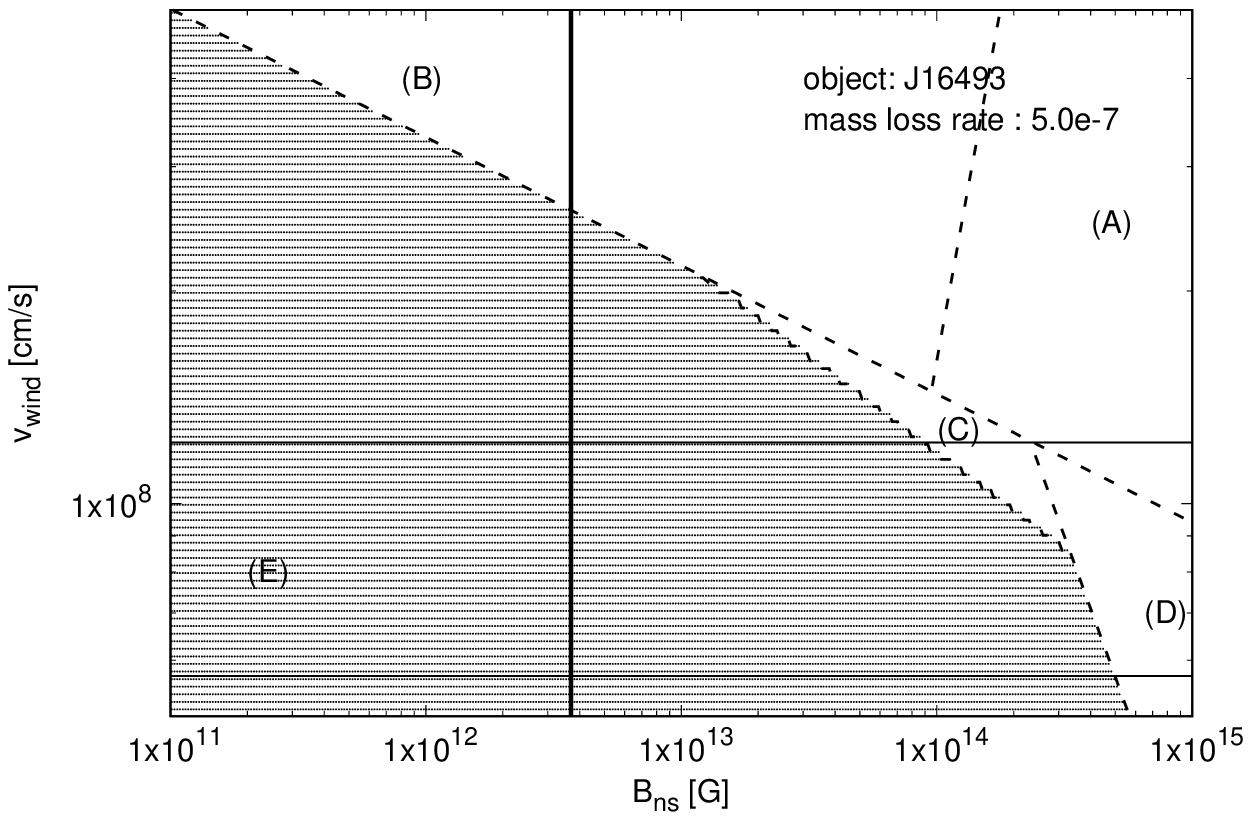} \\
\includegraphics[width=4.7cm]{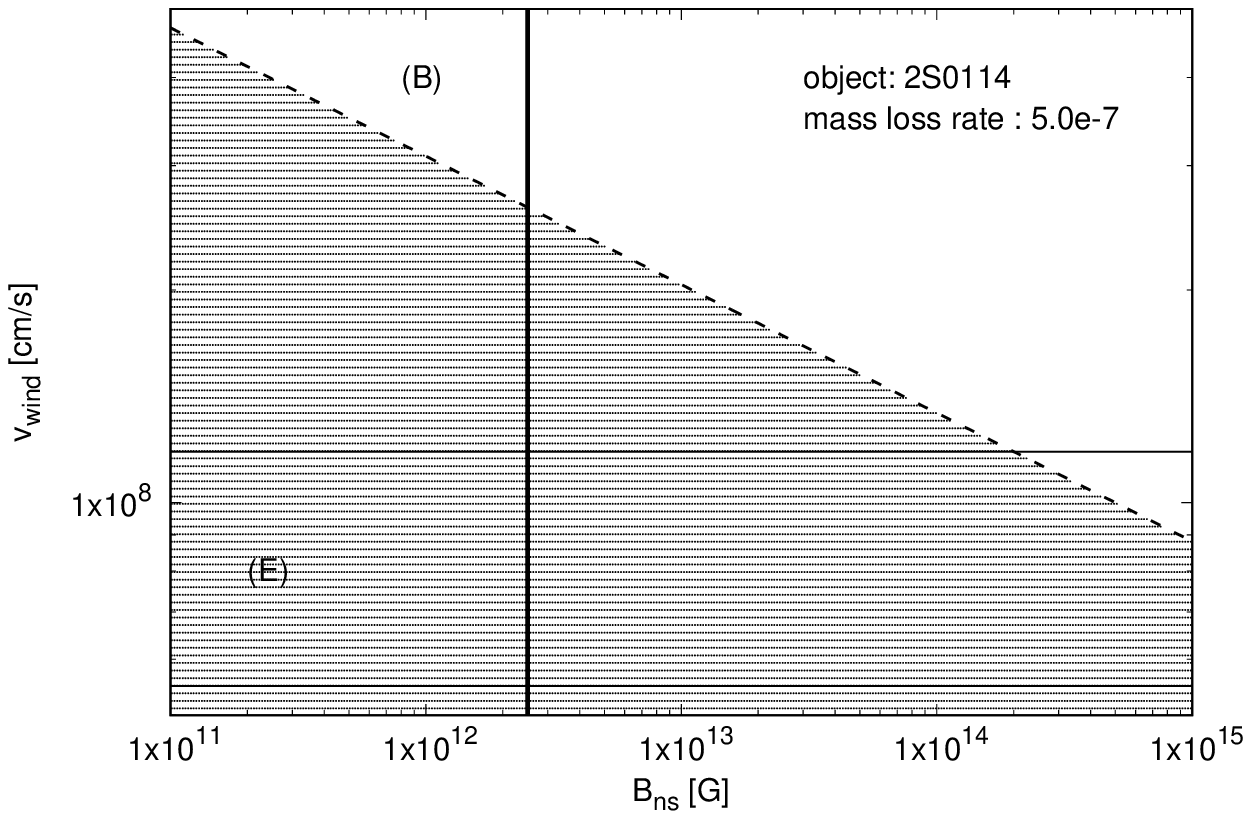}
\includegraphics[width=4.7cm]{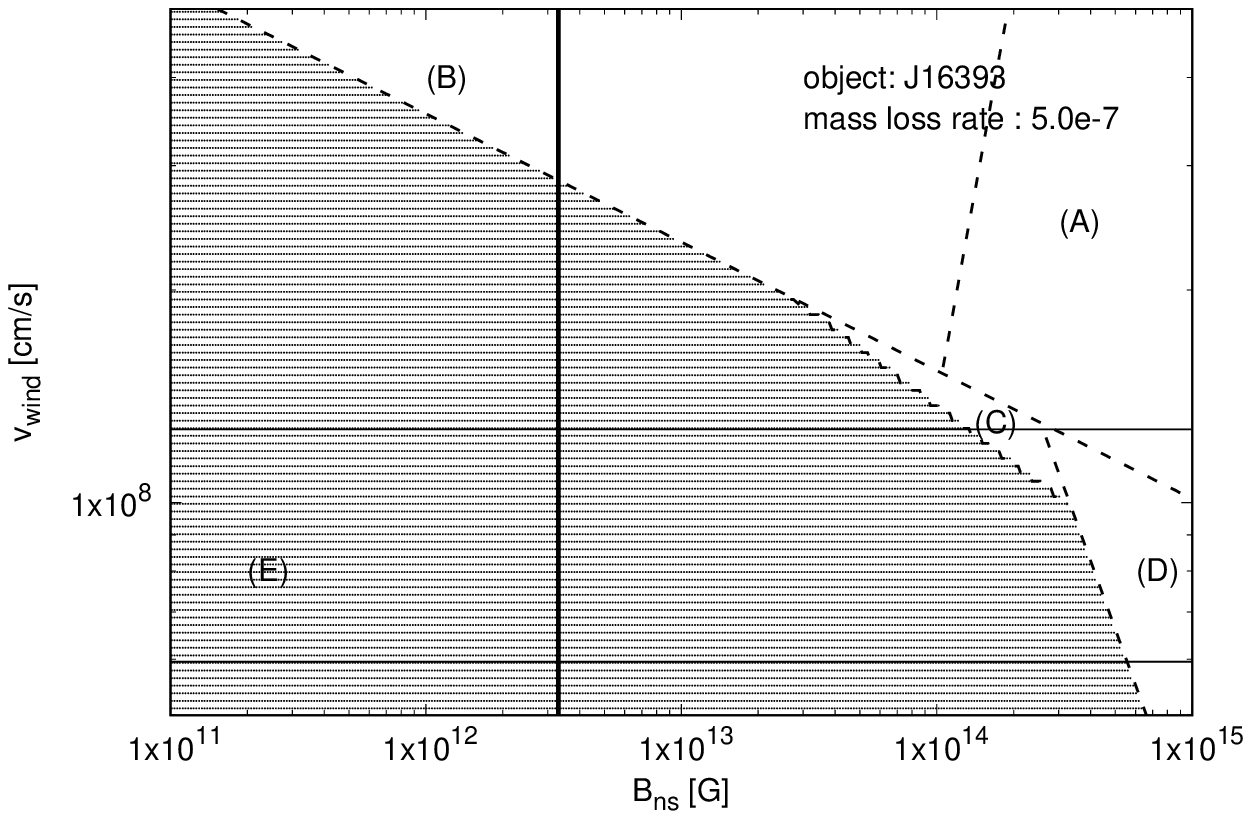}
\includegraphics[width=4.7cm]{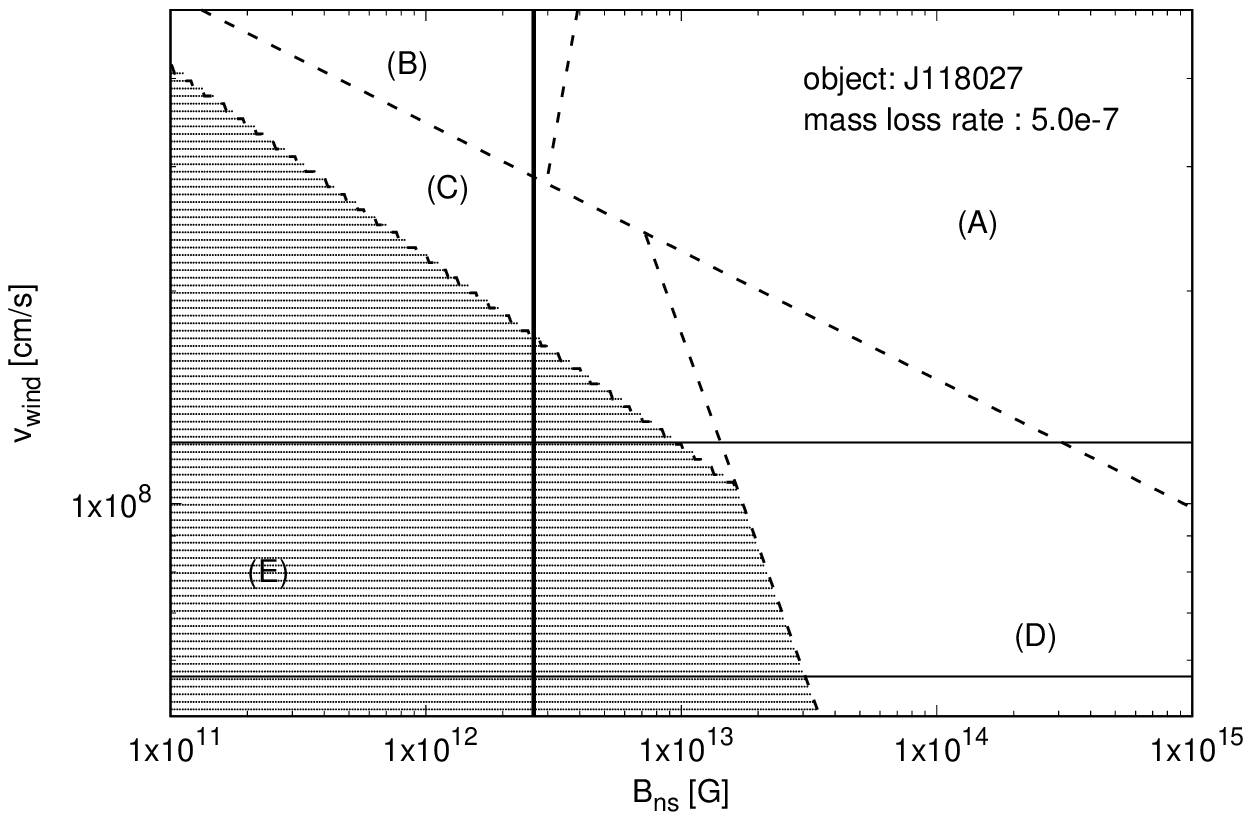}
\caption{
The accretion regime in the wind velocity-mass loss rate space for different systems.
The shaded region labeled as (E) denotes the direct accretion regime where a HMXB can be
observed as a bright X-ray source.
The two horizontal lines denote high ($1,500 \rm{km \,s}^{-1}$) and low ($700 \rm{km \,s}^{-1}$)
terminal velocity: $v_{\infty}$.
The vertical line shows the derived magnetic field from cyclotron lines.
If the region is being bounded by the vertical line with two horizontal lines (in the shaded region),
the system can be understood with the standard wind-fed scenario of SG-HMXB.
In our sample set, only two systems (LMC X-4 and OAO1657) cannot be explained with the standard model.}
\end{center}
\end{figure}

\begin{figure}
\begin{center}
\includegraphics[width=7cm]{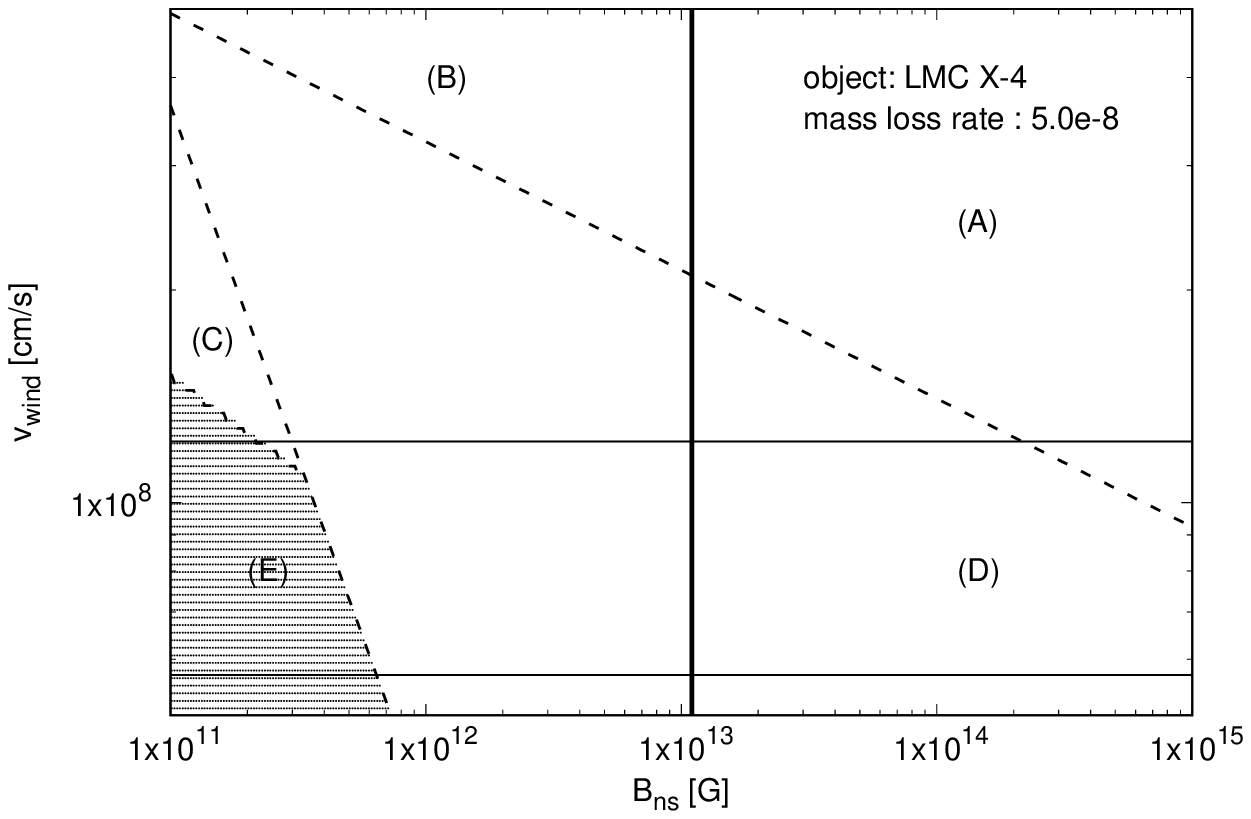}
\includegraphics[width=7cm]{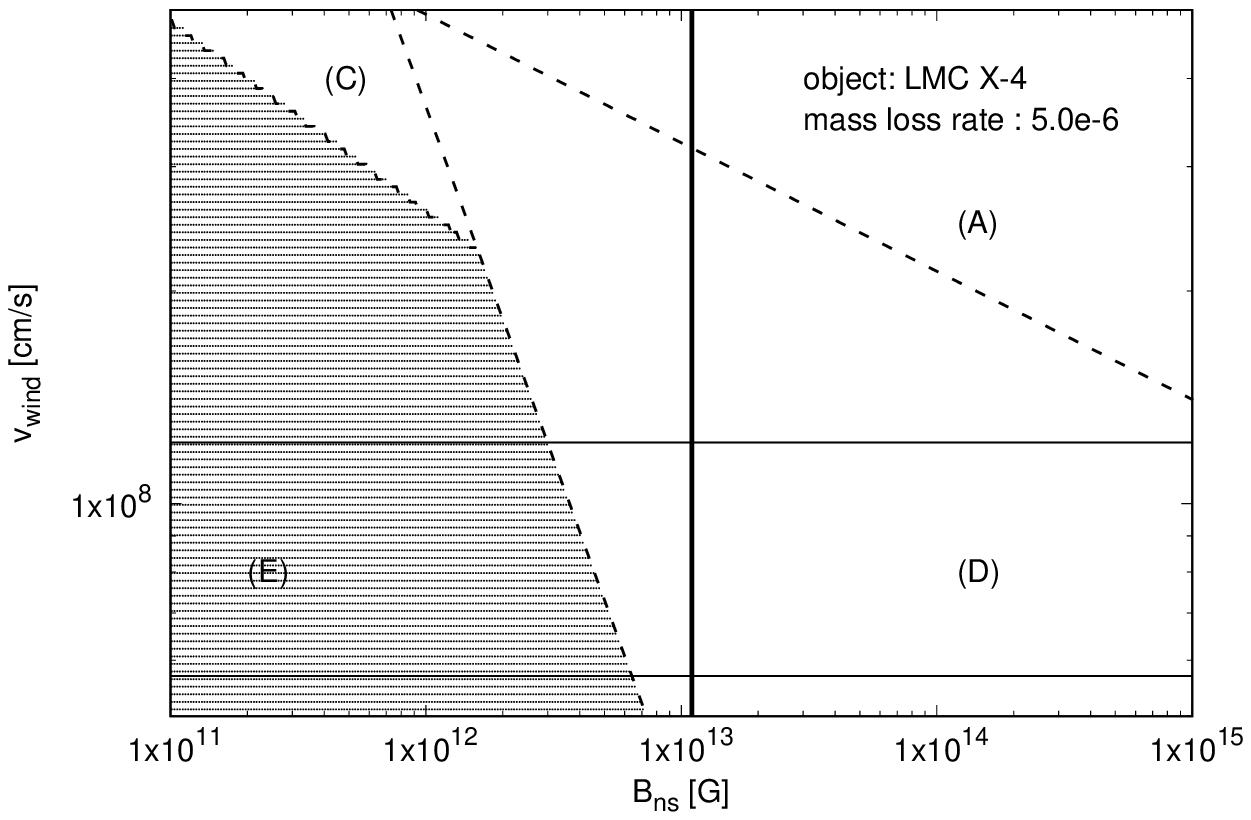}
\caption{
The same as Fig.~1, for LMC X-4 in the cases of high and low mass-loss rate:
$5 \times 10^{-8} M_{\odot} \rm{y}^{-1}$ (left panel) and
$5 \times 10^{-8} M_{\odot} \rm{y}^{-1}$ (right panel).}

\end{center}
\end{figure}

\begin{figure}
\begin{center}
\includegraphics[width=7cm]{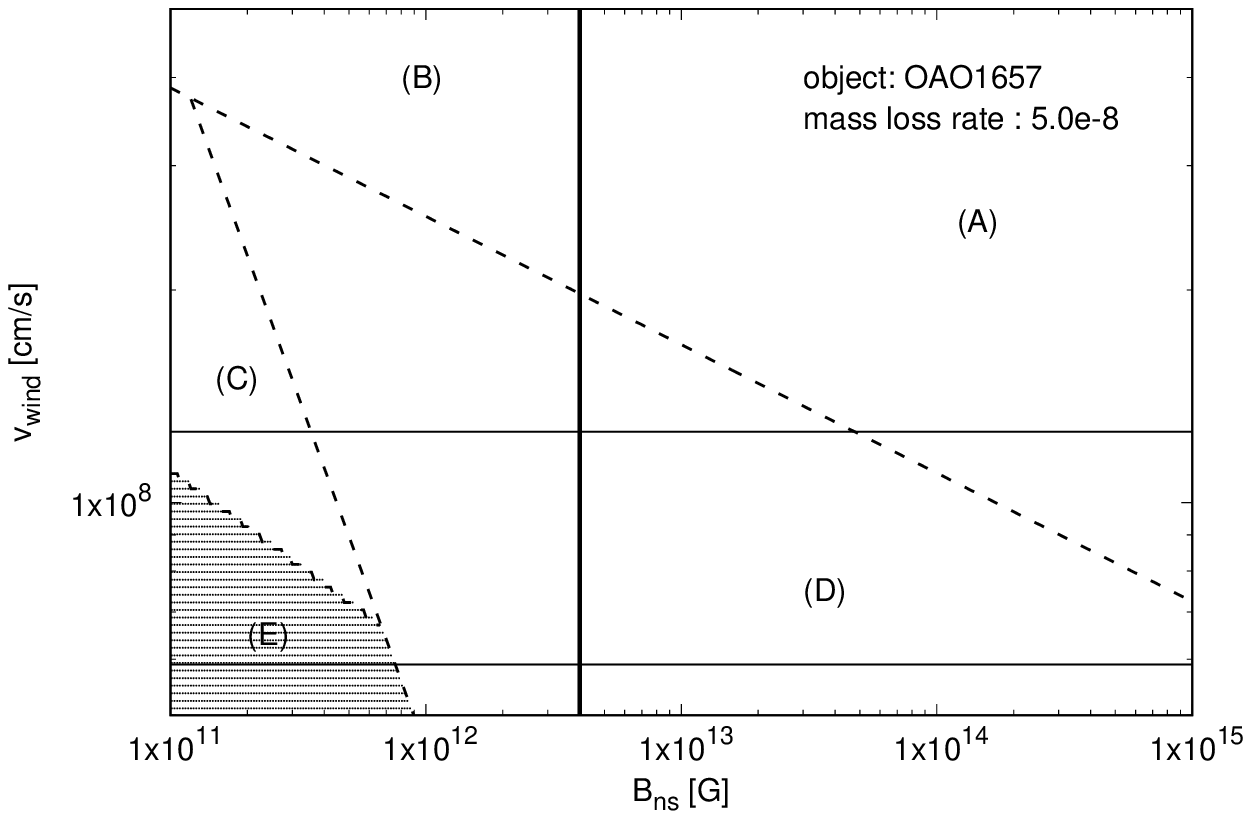}
\includegraphics[width=7cm]{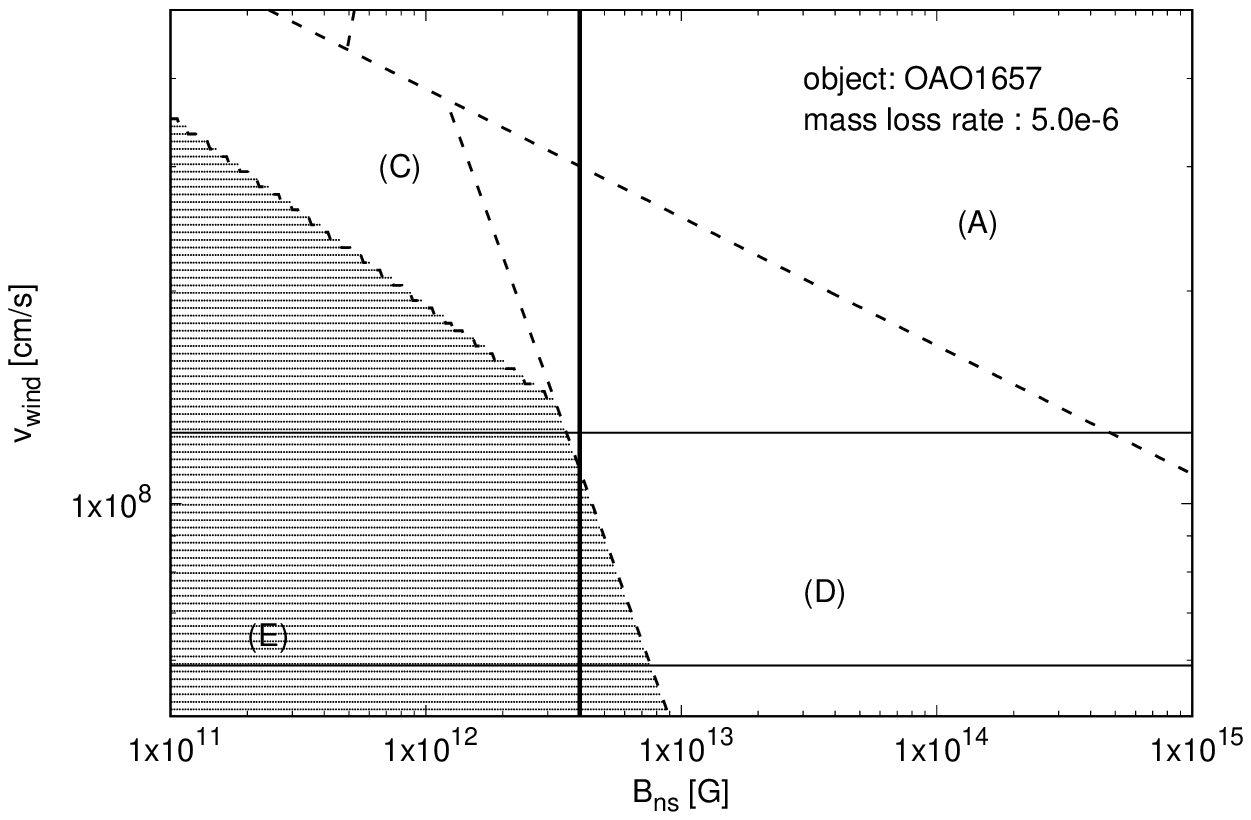}
\caption{
The same as Fig.~2, for OAO1657 in the cases of high and low mass-loss rate.}
\end{center}
\end{figure}

\section{Discussions}

Our method based on the equations gives a relationship between  $v_{\rm{w}}$ as a function of $B_{\rm{NS}}$.
We plotted  figures space (Figs. 1-3)  for all our sample.
We have considered two extreme cases of wind velocity in these figures ($1,500 \rm{km \, s}^{-1}$ for fast wind
and $700 \rm{km \, s}^{-1}$ for slow wind).
The lower sequence is rather realistic one for persistent wind-fed HMXBs, and the upper bound is rather typical for SFXTs
(Gim\'{e}nez-Garc\'{\i}a et al. 2016).
The direct accretion region is also plotted in the same figures, which qualitatively explain their different X-ray behavior.
From these figures (especially Figs.~2 and 3), it has been shown that a smaller $\dot{M}_{\rm{w}}$ could be rejected.

Another important parameter of HMXB which we have not examine in this study is the variation of the spin.
The numerical calculations done by many authors  suggest that there is no significant influence of angular momentum
transfer onto the NS from the wind of SG sources (Ruffert 1999; Fryxell \& Taam 1988; Matsuda et al. 1992;
Anzer \& B\"{o}rner 1995; Reig et al. 2009;  Ikhsanov \& Mereghetti 2015).
Even in slow wind velocity, the flow structure seems to be insufficient to exert a spin-up torque onto the NS.
Since the transfer of angular momentum during the accretion phase is  in-efficient.
This could show a small deviation from the instantaneous  spin periods at the early stage of the accretion
phase (Dai et al. 2016).

We have to notice here that if the accretion rate is too small $\dot{M} \geq 10 ^{8-10} M_{\odot} \rm {yr}^{-1}$,
the shock-crossing time would become so long and the angular momentum could be diffused during
passing the shock (see Shakura et al. 2012).
If enough angular momentum is transfered around the NS, an accretion disk may be formed outside the magnetic
radius under certain conditions.
While, at the higher accretion rate, the disk could not be a standard disk and expands, makes rather a torus.
In our preliminary analysis, the mass loss rate from the donor should be
between $10^{-7}$ to $10^{-5}$ $M_{\odot} \rm {yr}^{-1}$ to form an accretion disk.
This range seems to be reasonable from a standard mass loss theory of supergiant stars.
Therefore, the magnetic field is responsible for the formation of an accretion column since it forces the particles to
hit the NS surface at its magnetic poles.
In principle, more complex magnetic fields than a dipole field are possible.
It is noteworthy to mention here that, the relatively high X-ray luminosity such as J16493 makes these sources
the prime candidates for being disk-fed systems (see Reig et al. 2009; Falanga et al. 2015).
Whereas rather dim systems are all thought to be wind-fed systems (Shakura \& Postnov 2017).

It may be interesting to investigate the relation between the wind velocity as a function of spin period.
These parameters greatly affect the model of wind-fed binary systems and can be constrained during the
binary evolution.
It is clear that, the  sudden changes in the wind density as well as the mass lose rate, may lead to
the switching from one accretion regime to the other.
It should be noted that, the eccentricity of the orbit ($e \leq 0.25$) would lead also to additional variations
in the orbital separation (and consequently in $v_{\rm{w}}$), which reinforce the intrinsic variability of the stellar wind
and its capability to lead to transitions across regimes (Shenar 2016).

It is convenient to separate the evolution of the sources based on their accretion regimes as well as the stellar parameters.
The unusual properties (slow rotator, low X-ray luminosity and a super-orbital periodicity of 30.7 days (Farrell et al., 2006))
like in 1A 0114+650 would be important to get worth investigating about the origin of the X-ray behavior.
It suggests that this source evolves on the time scale of several years (Wang 2010), or it was born as a magnetar with
$B \sim 10^{14} \rm{G}$ (Sanjurjo-Ferrrin et al. 2017; Tong \& Wang 2018).
However, the propeller effect can spin down the NS to 5$ \times 10^{3}$ s (Reig et al. 2009) keeping the same values
as above but taking the magnetic field $10^{14} \rm{G}$ two orders of magnitude higher.
On the other hand a note should be made concerning the systems with slow wind (such as OAO1657-415).
This source is considered a suitable candidate to be a wind-fed system for most of the time among the known HMXBs,
and it is undergoing an episode of enhanced accretion from a temporary disk (Falanga et al. 2015).
Since this binary system is a wide range for Roche-lobe overflow to occur.
This may provide the clear evidence that winds in HMXBs possess sufficient angular momentum to form accretion disks
(see, e.g., Chakrabarty et al. 1993; Bildsten et al. 1997).
On contrary, in tight systems (such as LMC X-4), the donor does not fill its Roche lobe (see Falanga et al. 2015).
As explained before, this would imply the possibility that even in wind-fed systems, in a certain situation,
enough amount of angular momentum could be transferred via stellar wind and accretion matter may form an accretion disk.

\section{Conclusions}



We have performed parameter search computations to study the physical parameters of wind terminal velocity, mass loss rate,
and then the magnetic field of HMXBs with
supergiant companions. These quantities are important aspects of their
evolution due to the interaction between the components of the binary system.
We have also studied  the accretion regimes modeled by Bozzo et al. (2008) and prospects corresponding to the mass loss rate.
Depending on their theory, this will allow us to plot the v$_{wind}$-B$_{field}$ diagram
of several sources at different mass-lose rates, where the different accretion regimes occupied by different
space of parameters. This would help us to study their evolutionary scenario.

We show that 6 of 9 systems show good agreement with the standard wind-fed scenario.
they are 4U1907, 4U1538, J16493, 2S0114, J16393 and J18027.
Additionally, Vela X-1 satisfies our constraint in $v_{wind} - B_{field}$ plane with $\dot{M} \leq 5 \times 10^{-7}M_{\odot} \rm{y}^{-1}$.
We assume that the mass-loss rate is reasonably high ($\dot{M} \sim 10^{-6} M_{\odot} \rm{y}^{-1}$). As a result, the wind velocity
would be limited below $1,000 \rm{km \, s}^{-1}$. This result is consistent with recent result given by Gim\'{e}nez-Garc\'{\i}a et al. (2016).
On the other hand, 2 systems (LMC X-4 and OAO1657) cannot satisfy the wind-fed condition, since the reasonable band-region of $v_{\infty}$ cannot cross with the observed B-$\rm _{field}$ in the direct accretion regime.

Our study shows that only when the system enters the inner edge of accretion regime, it can emit bright X-rays and can be observed as a SG-HMXB. Characterized by the zone of cross-section of the magnetic filed in the figures (vertical line) with the wind velocity given by equations (two horizontal lines).

As has been correctly pointed out, several sources show variations  in wind velocity in the range of about 500 - 2,300 $\rm{km \, s}^{-1}$.
These variations  can be understood during the rotation phase at different parts, and also due to change in structure of magnetic field through to accretion dynamics, e.g. change in accretion rate, as is seen for sources like Vela X-1, J118027 and OAO1657.

The mass loss is proceed via spherical wind and the terminal velocity can be obtained in well-defined way for a number of wind-fed sources, where the donor is supergiant-type. According to our model, an enough amount of the angular momentum can be transferred via stellar wind and accreted  matter to form an accretion disk. This would make it happened in tight systems such as in J18027-201, J16393-4643  and 4U 1538-52. Or in systems with slow wind compared to the relative orbital velocity of the system, such as OAO 1657-415 and 2S 0114+65. As a result, more interaction between the accreted matter and the stars occurs, which may turn the binary into a good HMXB candidates via wind fed accretion around NS. One can relate the changes in accretion rates with the formation of a temporary accretion disk.

Furthermore, we have found that the range between  $10^{-7}$ to $10^{-5}$ $M_{\odot} \rm {yr}^{-1}$ represents the critical value of the disk to be
formed in the wind-fed X-ray binary systems. This range seems to be reasonable from a standard mass loss theory of
supergiant systems. As a result, if the wind mass-loss rate decreased,
this would lead to the disruption of the disk  and disappearance of the X-ray emission.
Which supports the reliability of current mass-loss rate predictions.

Finally, however, our model explains several important aspects of the behavior of disk-fed accreting X-ray binaries from low-level stellar wind accretion. In a few sources, the wind-accretion scenario cannot be applied. Thus, other scenarios (i.e. disk formation, partial Roche
lobe overflow or quasi Roche lobe) should be considered for further evolution for such binaries.




%
%
%

%

%

\section*{Acknowledgements}
We would like to thank the anonymous referee for the very
helpful comments in order to improve this paper. A. Taani would like to thank Abdul Hamed Shoman Founadation  for supporting this project (grant number 6/2017).
A. Taani grateful acknowledges hospitality from the Institute of High Energy Physics, Chinese Academy of Sciences through the
CAS President's International Fellowship Initiative (PIFI) 2018.
We are grateful to Motoki Nakajima and  Jiri Krti\v{c}ka comments and suggestions that allowed to improve the clarity of the
original version. SK is supported by JSPS KAKENHI grant 18K03706. L. Song  is supported by National Key R\&D Program of China (2016YFA0400801) and NSFC grant U1838201.


%
%


\begin{thebibliography}{999}


\bibitem{Aba15}
Abate C., Pols O. R., Karakas A. I., \& Izzard R. G. 2015, A\&A, 576, A118

\bibitem{Abu04}
Abubekerov M. K., 2004, Astron. Rep., 48, 649

\bibitem{Anz95}
Anzer U. \& B\"{o}rner G., 1995, A\&A, 299, 62A

\bibitem{Bha91}
Bhattacharya D. \& van den Heuvel E. J., 1991, Phys. Rep. 203, 1

\bibitem{Bil97}
Bildsten L., Chakrabarty D., Chiu J. \& et al. 1997, ApJS, 113, 367

\bibitem{Bof14}
Boffin H. M. J., 2014, Astrophysics and Space Science Library (Springer)




\bibitem{Bon44}
Bondi H., Hoyle F., 1944, MNRAS, 104, 273


\bibitem{Boz08}
Bozzo E., Falanga M. \& Stella L., 2008, ApJ, 683, 1031

\bibitem{Cas75}
Castor J. I.,  Abbott D. C. \& Klein R. I., 1975, ApJ, 195, 157

\bibitem{Cha93}
Chakrabarty D., Grunsfeld J. M., Prince T. A. \& et al., 1993, ApJ, 403, L33



%
\bibitem{Cob02}
Coburn W.,  Heind W. A.,  Rothschild R. E. \& et al. 2002, ApJ, 580,
394
\bibitem{Cla02}
Clark G. W., Woo J. W., Nagase F. \& et al. 1990, ApJ, 353, 274

\bibitem{Col15}
Coley J.B., Corbet R.H.D., \& Krimm H. A., 2011, ApJ, 808, 140


\bibitem{Cus98}
Cusumano G., de Salvo T., Burderi L. \& et al.  1998, A\&A, 338, L79

\bibitem{Dai06}
Dai H.-L. \& Li  X.-D., 2006, A\&A, 451, 581
\bibitem{Dai16}
Dai H.-L., Liu X.-W. \& Li  X.-D., 2016, MNRAS, 457, 3889D

\bibitem{Dai17}
Dai Z.B.,  Szkody P., Taani A. \& et al. 2017, A\&A, 606A, 45D
%
%
%
%
%
%
%
%
%

\bibitem{Far06}
Farrell S. A., Sood R. K., O'neill P. M., 2006, 
MNRAS,  367, 1457
%
\bibitem{Fal15}
Falanga  M.,  Bozzo  E.,  Lutovinov  A. \& et al. 2015, A\& A, 577, A130

%
%
%
%
\bibitem{Fel03}
Feldmeier A., Oskinova L. \& Hamann W. R., 2003, A\&A, 403, 217

\bibitem{Fry88}
Fryxell B.A. \& Taam R.E., 1988,  Astrophys. J. 335, 862
%
%
%
%
%
%
%
%
%
\bibitem{Ikh15}
Ikhsanov N. R. \& Mereghetti S., 2015, MNRAS, 454, 3760I
%
%
%
%
  \bibitem{Kar16}
 Karino S. \& Miller J. C. , 2016, MNRAS, 462, 3476K
 \bibitem{Kar18}
Karino K., Nakamura K. \&  Taani A., 2018, in preperation
\bibitem{Klu0705}
Klu\'{z}niak W., \& Rappaport S., 2007, ApJ, 671, 1990
 \bibitem{Krt15}
 Krti\v{c}ka J., Kub\'{a}t J., \& Krti\v{c}kov\'{a} I. 2015, A\&A, 579, A111
  \bibitem{Krt16}
 Krti\v{c}ka J., Kub\'{a}t J., \& Krti\v{c}kov\'{a} I. 2016, A\&A, 593A, 101K
  \bibitem{Krt17}
 Krti\v{c}ka J., \& Kub\'{a}t J., 2017, A\&A, 606A, 31K
  \bibitem{Kud00}
  Kudritzki R.-P., \& Puls, J. 2000, ARA\&A, 38, 613
%
%
%
%
%
\bibitem{Gim16}
Gim\'{e}nez-Garc\'{\i}a A., Shenar T., Torrej\'{o}n J. M. et al. 2016, A\&A, 591A, 26G
%
%
 \bibitem{Lut17}
 Lutovinov, A., Tsygankov, S. 2017, MNRAS, 466
\bibitem{Lam99}
Lamers H. J., Haser S., de Koter A., \& Leitherer C., 1999, ApJ, 516, 872
\bibitem{Liu06}
Liu Q. Z., van Paradijs J., \& van den Heuvel E. P. J. 2006, A\&A, 455, 1165
\bibitem{Mac91}
MacFarlane J. J., Cassinelli J. P., Welsh, B. Y. \& et al. 1991, ApJ, 380, 564
\bibitem{Mak90}
Makishima K., Mihara T., Ishida M. \& et al. 1990, ApJ 365, L59
%

%
%
%
\bibitem{Man12}
Manousakis A., Walter \& J. M. Blondin J. M., 2012, A\&A, 547, A20

\bibitem{Mas12}
Mason A. B., Clark J. S., Norton A. J. \& et al. 2012, MNRAS, 422, 199

\bibitem{Mat92}
Matsuda T., Ishii T., Sekino N. \& et al., 1992, MNRAS, 255, 183
\bibitem{Mus15}
Mushtukov A., Tsygankov S.,  Serber A. \& et al. 2015, MNRAS, 454, 2714M

\bibitem{Naj11}
Najarro F., Hanson M. M., \& Puls J., 2011, A\&A, 535, A32

\bibitem{Nag04}
Nagae T., Oka K., Matsuda T. \& et al. 2004, A\&A, 419, 335
%
%
%
%


\bibitem{Nes10}
Nespoli E., Fabregat J., \&  Mennickent R. E., 2010, A\&A, 516, A106

\bibitem{Nish03}
Nishimura O. 2003, PASJ, 55, 849

\bibitem{Nish11}
Nishimura O., 2011, ApJ, 730, 106

\bibitem{Pri}
Pringle J. E., \& Rees, M. J., 1972, A\&A, 21, 1

\bibitem{Pos17}
Postnov K., Oskinova L., \& Torrej\'{o}n J., 2017, arXiv; 170100336P

\bibitem{Pot12}
Pottschmidt K., Suchy S., Rivers E. \& et al. 2012, AIPC, 1427, 60P

%
%

\bibitem{Pul06}
Puls J., Markova N., Scuderi S., \& et al. 2006, A\&A, 454, 625
%
%
%
%
\bibitem{Raw11}
Rawls M. L., Orosz J. A., McClintock J. E. \& et al. 2011, ApJ, 730, 25
\bibitem{Rei09}
Reig P., Torrej\'{o}n, J. M., Negueruela I. \& et al., 2009, A\&A, 367, 266
\bibitem{Rei16}
Reig P., Nersesian A., Zezas A. \& et al. 2016, A\&A, 590, A122

\bibitem{Rob01}
Robba N. R., Burderi L., Di Salvo T. \& et al.  2001, ApJ, 562, 950

\bibitem{Rod09}
Rodes-Roca J. J., Torrej\'{o}n J. M., Kreykenbohm I. \& et al. 2009,
A\&A, 508, 395
%
\bibitem{Ruf99}
Ruffert M., 1999,  A\&A, 346, 861
%
%
%
%
%
%
%
	
\bibitem{Sal18}
Saladino M. I., Pols O. R., van der Helm E. \& et al. 2018, arXiv: 180503208

\bibitem{San17}
Sanjurjo-Ferrrin G., Torrejon J. M., Postnov K., et al., 2017, A\&A, 606, A145

\bibitem{Sha12}
Shakura N., Postnov K., Kochetkova A. \& Hjalmarsdotter L.,  2012, MNRAS, 420, 216

\bibitem{Sha17}
Shakura N. \& Postnov K., 2017, arXiv: 170203393

\bibitem{She16}
Shi C.-S., Zhang S.-N., \& Li X.-D., 2015, ApJ, 813, 91S

\bibitem{She16}
Shenar T., 2016, PhD thesis, University of Potsdam,


\bibitem{sta07}
Staubert R., Shakura N., Postnov K. \& et al. 2007, A\&A, 465, 25S

\bibitem{ste86}
Stella L., White N. E. \& Rosner, R. 1986, ApJ, 308, 669

\bibitem{Taa12a}
Taani A.,  Zhang C.M., Al-Wardat M.,  \& Zhao Y. H.,  2012a, 
AN, 333, 53

\bibitem{Taa12b}
Taani A.,  Zhang C. M., Al-Wardat M., \& Zhao Y. H., 2012b, 
Ap\&SS, 338, 295

\bibitem{Taa16}
Taani A., 2016, RAA, 16, 101


\bibitem{Taa17}
Taani A. \&  Khasawaneh A.,  J. of Phy. Conf. Ser., 2017, 869,  012090


\bibitem{Taa18}
Taani A.,  Karino K., Song L. \& et al. 2018, arXiv: 180805345T

\bibitem{Tom06}
Thompson T. W., Tomsick J. A., Rothschild R. E., \& et al. 2006, ApJ, 649, 373

\bibitem{Tru78}
Tr\"{u}emper J., Pietsch W., Reppin C. \& et al. 1978, ApJ, 219, 105

\bibitem{Ton15}
Tong H., 2015, RAA, 15, 517

\bibitem{Ton18}
Tong H. \& Wang W., 2018, arXiv:180605784T

\bibitem{Van04}
van den Heuvel E. P. J. 2004, Science, 303, 1143

\bibitem{Vin01}
Vink J. S., de Koter A. \& Lamers H. J., 2001, A \& A, 369, 574

\bibitem{Vog82}
Voges W., Pietsch W., Reppin C. \& et al., ApJ, 1982, 263, 803


\bibitem{Wal15}
Walter R., Lutovinov A. \& Bozzo E., 2015, A\&ARv, 23, 2W

\bibitem{Wanl08}
Wang Y.-M. \& Robertson J. A., 1985, A\&A, 151, 361

\bibitem{Wan95}
Wang Y.-M., 1995, ApJ, 449, L153


\bibitem{Wil08}
Wilson C., Finger M. \& Camero-Arranz A., 2008,  ApJ, 678, 1263
\bibitem{wei10}
Wei Y. C., Taani A., Pan Y. Y. et al.,  Chin. Phys. Lett., 2010, 27, 9801

\bibitem{Zha06}
Zhang C. M. \& Kijima Y., 2006, MNRAS, 366, 137


\end{thebibliography}
\end{document}